\documentclass[12pt]{article}
\usepackage{epsfig}

\def\beq{\begin{equation}}
\def\eeq#1{\label{#1}\end{equation}}
\def\eeqn{\end{equation}}

\def\beqa{\begin{eqnarray}}
\def\eeqa#1{\label{#1}\end{eqnarray}}
\def\eeqan{\end{eqnarray}}

\let\bar=\overbar

\def\Dslash{\not{\hbox{\kern-4pt $D$}}}
\def\dslash{\not{\hbox{\kern-2pt $\del$}}}

\def\msb{{\bar{\ssstyle M \kern -1pt S}}}

\def\Title#1{\begin{center} {\Large {\bf #1} } \end{center}}

\begin{document}

\Title{Searches in CMS}

\bigskip\bigskip


\begin{raggedright}  

{\it Lars Sonnenschein\index{Sonnenschein, L.}  \hfill on behalf of the CMS Collaboration \\
RWTH Aachen University \\ 
III. Phys. Inst. A  \hfill Submitted to proceedings of PLHC 2011 \\
D-52056 Aachen, GERMANY 
}
\bigskip\bigskip
\end{raggedright}

\section{Introduction}

Searches for physics beyond the Standard Model (SM) with the CMS experiment~\cite{cms}
in $pp$ collisions at a centre of mass energy of $\sqrt{s}=7$~TeV at the LHC are presented. 
The discussed results are based on datasets of 2010 (34 - 40 pb$^{-1}$) and 2011 (190 - 191 pb$^{-1}$).
A complete list of public analyses can be found at ref. \cite{SusyPub} for $R$-parity conserving 
super symmetric searches and at ref. \cite{ExoPub} for exotic searches. 
Various important theories, encompassing new heavy resonances, quark/lepton compositeness, extra 
dimensions as well as other exotic signatures are tested.
In section \ref{susysearches} 
a few supersymmetric searches are shown.
Several other supersymmetric hadronic analyses \cite{sus-10-005} \cite{sus-10-009} that have been 
recently submitted extend the exclusion limits significantly.
The following section \ref{gravitysearches} covers searches for TeV scale gravity.
The remainder of presented exotic signal searches is subdivided according to their final states
into three categories: lepton production (section \ref{leptonprod}), lepton + jets production 
(section \ref{lepjetsprod}) and jet production (section \ref{jetsprod}).   
In general the different search channels set also model independent exclusion limits on the production 
of events above the SM expectation. Jets are reconstructed from particle flow objects (if not indicated
otherwise) by means of the anti-$k_T$ jet algorithm using a distance parameter of $R=0.5$.

\section{Supersymmetric searches} \label{susysearches}

\subsection{Search in opposite sign dilepton events}
This search \cite{sus-10-007} makes use of an integrated luminosity of 34~pb$^{-1}$ and 
extends existing Tevatron and LEP exclusion limits.
Events with two oppositely charged leptons accompanied by at least two hadronic jets are considered.
Furthermore requirements on the scalar sum of selected jet transverse energies $H_T$ and 
on the missing transverse energy $E\!\!\!\!/\,_T$  are made. The background in the signal region is
determined by means of data driven techniques based on four regions of the almost uncorrelated variables
$H_T$ and $y := E\!\!\!\!/\,_T / \sqrt{H_T}$ and alternatively based on the dilepton transverse momentum
($p_T(\ell\ell)$) method, which models the $E\!\!\!\!/\,_T$ in considering the leptons as neutrinos.
No significant excess over the SM expectation is observed.
The Bayesian 95\% Confidence Level (CL) upper limit on the number of non-SM events in the signal
region is 4.0 ($e\mu$) and 3.0 ($ee, \mu\mu$).
These limits can be interpreted in the Constrained Minimal Supersymmetric SM (CMSSM). The CMS
Low Mass benchmark points LM0 and LM1 defined in the mSUGRA model in the plane of the gaugino mass
$m_{1/2}$ and the scalar mass $m_{0}$ (for $\tan\beta=3$, $A_0=0$, $\mu>0$) are already excluded at 95\%
CL.

\subsection{Search in dijet + $E\!\!\!\!/\,_T$ events}
This search \cite{sus-10-003} makes use of an integrated luminosity of 35 pb$^{-1}$.
It extends existing Tevatron and LEP exclusion limits.
Calorimeter towers instead of particle flow objects are clustered by the anti-$k_T$ jet algorithm 
($R=0.5$) for trigger efficiency reasons wrt the considered data taking period.
Events are selected with at least two jets, $H_T$ and $H\!\!\!\!/\,_T$, defined as the vector sum 
of jet transverse momenta, excluding the jet which maximises its azimuthal angle to $H\!\!\!\!/\,_T$.   
Events with isolated leptons, photons and relatively large calorimetric $E\!\!\!\!/\,_T$ with respect 
to $H\!\!\!\!/\,_T$ are vetoed.
A variable $\alpha_T$, defined as the transverse energy of the second leading jet over the 
transverse mass of the dijet system is exploited to discriminate multijet background.
For higher jet multiplicities the jets are combined into two pseudo-jets such that their scalar 
transverse energy difference is minimised. 
A data driven method, based on ratios of the $\alpha_T$ variable in neighboured $H_T$ intervals is used
to determine the background in the signal region. No significant excess over the SM expectation is 
observed. Model independent limits on signal yields
are derived with a profile likelihood based on the Feldman-Cousins approach. 
No dedicated optimisation for the interpretation within the CMSSM has been applied.
The benchmark points LM0 and LM1 are excluded at the 99.99\% and 99.2\% CL respectively.

\subsection{Search in $Z$ + jets + $E\!\!\!\!/\,_T$ events}
This search \cite{sus-11-012} makes use of an integrated luminosity of 191 pb$^{-1}$.
Events with two isolated oppositely charged leptons in a mass window around the $Z$ boson mass
and at least three jets are considered. The Jet $Z$ boson Balance (JZB), defined as the 
transverse momentum difference between the jets and the two leptons is used to discriminate background.
$t\bar{t}$ background is determined from data in the dilepton opposite-flavour channel.
No significant excess over the SM expectation is observed.  
Model independent limits are obtained via Bayesian inference making use of a profile likelihood.
95\% CL upper limits on the production of signal events are set to $14^{+7.6}_{-4.6}$ (JZB$>50$~GeV)
and $8.6^{+4.0}_{-3.5}$ (JZB$>100$~GeV). Within the CMSSM cross section times branching ratio
times acceptance ($\sigma\times\mbox{BR}\times{\cal{A}}$) limits of 0.040~pb and
0.043~pb are obtained at the benchmark points LM4 and LM8 respectively.

\section{Search for TeV scale gravity} \label{gravitysearches}

\subsection{Search for Large Extra Dimensions in dimuon events}
The ADD model \cite{add} postulates the existence of additional compactified extra dimensions.
An effective Planck scale 
$M_D^{n+2} = M^2_{Pl} / (8\pi L^n)$
reduces the Planck scale $M_{Pl}$ depending on the number $n$ and size $L$ of Large Extra 
Dimensions (LED). This provides one way to solve the hierarchy problem.
This search \cite{exo-10-020} makes use of an integrated luminosity of 40 pb$^{-1}$.
Events with two isolated muons are selected. In the absence of significant excess 
Bayesian upper limits at 95\% CL on a signal cross section of $\sigma = 0.088 - 0.098$~pb are set for 
invariant dimuon masses in the range of $1<M_{\mu\mu}<7$~TeV$/c^2$. These limits can be translated
into limits on an effective Planck scale up to 2.15~TeV at 95\% CL.

\subsection{Search for microscopic black hole signatures}
This search \cite{exo-11-021} makes use of an integrated luminosity of 190 pb$^{-1}$.
Events with enhanced scalar sum $S_T$ of transverse energies of electrons, photons, muons, jets
and $E\!\!\!\!/\,_T$ are considered. The variable $S_T$ has a reduced sensitivity to initial
and final state radiation. The analysis is accomplished separately for different number of
reconstructed objects $N(e,\gamma,\mu,\mbox{jet})=2,3,4,5,6$.
No significant excess over the SM prediction has been observed.
Model independent exclusion limits at 95\% CL are set on the cross section times acceptance
of order 15~fb$^{-1}$. Black hole masses of 3.8 - 4.9~TeV are excluded in dependence of 
the effective Planck scale of the ADD model \cite{add} whose exclusion limits vary between 
3.7 and 4.9~TeV.

\subsection{Search for Randall-Sundrum gravitons in diphoton events}
This search \cite{exo-10-019} makes use of an integrated luminosity of 36 pb$^{-1}$ and 
benefits from a factor of two enhanced signal cross section compared to graviton production
with a fermion pair in the final state.
Two isolated photons are required. 
No significant excess over the SM prediction has been observed.
95\% CL upper limits on $\sigma\times\mbox{BR}$ are 
set in different diphoton invariant mass windows and translated into limits on a graviton mass $M_1$
as a function of the coupling ratio $k/M_{Pl}$.

\subsection{Search in Mono-Jet + $E\!\!\!\!/\,_T$ events}
This search \cite{exo-11-003} makes use of an integrated luminosity of 36 pb$^{-1}$.
Events with at least two jets, no isolated leptons and $E\!\!\!\!/\,_T$ are selected.
No significant excess over the SM prediction has been observed.
Bayesian 95\% CL upper limits on number of non-SM events compatible with the measurement are set 
and translated into limits on model cross sections and parameters. 

\section{Searches in lepton production} \label{leptonprod}

\subsection{Search in highly boosted $Z\rightarrow\mu^+\mu^-$ events}
This search \cite{exo-10-025} makes use of an integrated luminosity of 36 pb$^{-1}$.
It is optimised for excited quarks, making a transition to a SM quark in radiating a $Z$ boson.
This analysis is sensitive to compositeness, supersymmetry, Technicolour and new gauge bosons.
A pair of charge conjugated isolated muons in a $Z$ boson mass window is required.
No significant excess over the SM prediction has been observed.
Bayesian 95\% CL upper limits for transverse dimuon momenta thresholds between 200 and 400~GeV
are set on excited quark masses $m_{q^*}$ assuming the mass equal to the scale $\Lambda$ and
SM like fermion couplings. Depending on the production mechanism excited quark masses 
$m_{q^*}$ around 1~TeV$/c^2$ are excluded.

\subsection{Search for resonant lepton jets}
This search \cite{exo-11-013} makes use of an integrated luminosity of 35 pb$^{-1}$
and extends the Tevatron reach significantly. It is sensitive to a broad range of models, 
including dark photons ($\gamma_{\mbox{\scriptsize dark}}\rightarrow\mu\mu$).
Events with muons collimated in muon-jets are selected. At least one charge conjugated muon pair
per muon-jet is required. The events are categorised depending on the number of muon-jets and
number of muons in each jet. 
No significant excess over the SM prediction has been observed.
Bayesian 95\% CL upper limits on new low mass states decaying into 
muon pairs are set for the different event categories. The limits on signal cross section times 
branching ratio time acceptance $\sigma\times\mbox{BR}\times{\cal{A}}$ range from 0.1 to 0.5~pb.

\subsection{Search in inclusive dilepton events}
This search \cite{exo-10-024} makes use of an integrated luminosity of 35 pb$^{-1}$ and is 
model independent. Events with isolated leptons and jets are selected. The events are separated
into dilepton invariant mass $Z$ and non-$Z$ regions. The signal region is defined by a more stringent
scalar transverse energy sum $S_T$ requirement.
No significant excess over the SM prediction has been observed.
95\% CL upper limits on signal production of $\sigma\times {\cal{A}}=0.14$~pb are set for both 
invariant dilepton mass regions.

\subsection{Search for excited leptons}
This search \cite{exo-10-016} makes use of an integrated luminosity of 36 pb$^{-1}$.
The excited lepton makes the transition to a SM lepton in radiating a photon in the
analysis channel considered here. Events with an isolated photon and an isolated pair of 
charge conjugated same-flavour leptons are selected. 
Bayesian 95\% CL upper limits
on the scale $\Lambda$ as a function of the excited lepton mass in the TeV range are set.

\section{Searches in lepton + jets production} \label{lepjetsprod}

\subsection{Search for 1st generation scalar leptoquarks } 
This search \cite{exo-10-006} makes use of an integrated luminosity of 36 pb$^{-1}$.
Events with exactly one electron - isolated - at least two jets, $E\!\!\!\!/\,_T$ and $S_T$
are selected. the signal region is defined by higher $S_T$ and transverse masses
$m_T^{e, E\!\!\!\!/\,_T}$ of the electron and $E\!\!\!\!/\,_T$.
No significant excess is observed.
Bayesian 95\% CL upper limits on leptoquark (LQ) pair production times branching ratio 
exclude $m_{\mbox{\scriptsize LQ}}>310$~GeV$/c^2$. Combination with the $eejj$ channel 
improve the exclusion limits correspondingly.

\subsection{Search for heavy bottom-like quarks}
This search \cite{exo-10-018} makes use of an integrated luminosity of 34 pb$^{-1}$.
Heavy bottom-like quark pair production 
$b'\bar{b}'\rightarrow tW^-\bar{t}W^+\rightarrow bW^+W^-\bar{b}W^-W^+$ is considered.
At least 2(3) isolated leptons and 4(2) jets are required. The signal region is defined by enhanced
$S_T$. Bayesian 95\% CL upper limits on the cross section as function of $m_b$ exclude $b'$ masses
between 255 and 361~GeV$/c^2$ are set.

\section{Searches in jet production} \label{jetsprod}

\subsection{Search for multijet resonances}
This search \cite{exo-11-001} makes use of an integrated luminosity of 35 pb$^{-1}$.
It is model independent but optimised for gluino pair production $pp\rightarrow\tilde{g}\tilde{g}$
and R-parity violating decay $\tilde{g}\rightarrow 3q$. 
Events with at least six jets and low invariant three jet masses are selected.
Bayesian 95\% CL upper limits on $\sigma\times\mbox{BR}$ are set, 
excluding $200<m_{\tilde{g}}<280$~GeV$/c^2$.

\subsection{Search for quark compositeness in dijet production}
This search \cite{exo-11-001} makes use of an integrated luminosity of 36 pb$^{-1}$.
The differential cross section 
is measured
as a function of $\chi_{\mbox{\scriptsize dijet}}$, defined as 
the exponential absolute dijet rapidity difference. 
Jets are determined from calorimeter towers. 
95\% CL limits in the modified frequentist approach ($CL_s$ method) exclude compositeness scales
up to $\Lambda^+ = 5.6$~TeV ($\Lambda^- = 6.7$~TeV) for destructive (constructive) interference. 

\section{Conclusions}
CMS is searching for evidence of new physics beyond the SM in many channels using early LHC
data at $\sqrt{s} = 7$~TeV. New territory beyond the Tevatron is already explored.
No signals of new physics have been observed in the early LHC data yet.

\end{document}